\documentclass{sig-alternate-2013}
\usepackage{graphicx}
\usepackage{algorithm}
\usepackage{algorithmic}
\usepackage{booktabs}
\usepackage{placeins}

\def\<{\langle}
\def\>{\rangle}

\newtheorem{proposition}{Proposition}[section]
\newtheorem{definition}{Definition}[section]
\newtheorem{theorem}{Theorem}[section]

\begin{document}
\title{Mining for Geographically Disperse Communities in Social Networks by Leveraging Distance Modularity}
\numberofauthors{3}
\author{
%
%
\alignauthor Paulo Shakarian\\
       \affaddr{Network Science Center and}\\
       \affaddr{Dept. of Electrical Engineering}\\
       \affaddr{and Computer Science}\\
       \affaddr{U.S. Military Academy}\\
       \affaddr{West Point, NY 10996}\\
       \email{paulo@shakarian.net}
\alignauthor Patrick Roos\\
       \affaddr{Dept. of Computer Science}\\
       \affaddr{University of Maryland}\\
       \affaddr{College Park, MD 20721}\\
       \email{roos@cs.umd.edu}
\alignauthor Devon Callahan,\\
Cory Kirk\\
       \affaddr{Network Science Center and}\\
       \affaddr{Dept. of Electrical Engineering}\\
       \affaddr{and Computer Science}\\
       \affaddr{U.S. Military Academy}\\
       \affaddr{West Point, NY 10996}\\
       \email{devon.callahan@usma.edu}\\
       \email{cory.kirk@usma.edu}
}
\maketitle
\begin{abstract}
Social networks where the actors occupy geospatial locations are prevalent in military, intelligence, and policing operations such as counter-terrorism, counter-insurgency, and combating organized crime.  These networks are often derived from a variety of intelligence sources.  The discovery of communities that are geographically disperse stems from the requirement to identify higher-level organizational structures, such as a logistics group that provides support to various geographically disperse terrorist cells.  We apply a variant of Newman-Girvan modularity to this problem known as distance modularity.  To address the problem of finding geographically disperse communities, we modify the well-known Louvain algorithm to find partitions of networks that provide near-optimal solutions to this quantity.  We apply this algorithm to numerous samples from two real-world social networks and a terrorism network data set whose nodes have associated geospatial locations.  Our experiments show this to be an effective approach and highlight various practical considerations when applying the algorithm to distance modularity maximization. Several military, intelligence, and law-enforcement organizations are working with us to further test and field software for this emerging application.
\end{abstract}
\category{Applied Computing}{Law, social and behavioral sciences}{Sociology}
\terms{Algorithms, Experimentation}

\keywords{complex networks, geospatial reasoning}

\section{Introduction}

In recent years, fueled by the connectivity of our social world and technological advances that allow for effortless collection of connectivity data, much effort has been invested in developing algorithms for the detection of communities in networks (e.g. \cite{girvan2002community,newman04,newman04a,du2007community,blondel08,Schaefer11,blondel11,cerina12}). The detection of communities - subsets of nodes that are highly connected in globally sparser networks - provides important insights into the organization of networks and related hidden information of social networks \cite{girvan2002community}. 

In many application domains, apart from the social network information provided by connectivity data, geospatial information is available as well, and community detection algorithms can be improved by leveraging such spatial information.
Social networks where the actors occupy geospatial locations are prevalent in military, intelligence, and policing operations such as counter-terrorism, counter-insurgency, and combating organized crime.  These networks are often derived from a variety of intelligence sources.
Community detection algorithms that specifically detect geographically dispersed communities are of interest in such application domains to identify higher-level organizational structures, such as a logistics group that provides support to various geographically disperse terrorist cells.  
Such communities may be less obvious in solely the available social network data. Hence, in order to find geographically dispersed communities, there exists a need for community detection algorithms that are optimized considering geospatial information in addition to social network information, and we address this need in this paper.
 

Blondel et al. \cite{blondel08} have developed a heuristic method known as the \textit{Louvain algorithm} that partitions a social network into communities while optimizing Newman-Girvan modularity of the partition. Newman-Girvan modularity is a common performance measure in community detection algorithms that gives a measure of how densely the detected communities of the partition are connected relative to connections between these communities \cite{newman04}. More specifically, the modularity measure is the ``fraction of edges within communities in the observed network minus the expected value of that fraction in a \textit{null model}, which serves as a reference and should characterize some features of the observed network'' \cite{liu12}.

In this paper, we use a variant of Newman-Girvan modularity with the Louvain algorithm to address the problem of mining for geographically dispersed communities in application domains where geospatial information is pertinent. Instead of the original null model used in Newman-Girvan modularity, we leverage a null model introduced by Liu et al. \cite{liu12}. The use of this model results in the \textit{distance modularity} measure of community structure. 

We test the algorithm on two real-world location-based social networks and a network from a transnational terrorism data set, the nodes of which have associated geospatial locations. Our experiments show that this approach is effective at finding partitions of networks that provide near-optimal solutions to distance modularity. We also highlight various practical considerations when applying the algorithm with these definitions of modularity. By testing the algorithm on a social network that is significantly larger (ca. 2100 nodes) than the test networks commonly used in the literature on community detection algorithms (typically $\lessapprox 600$ nodes), we also better demonstrate scalability. Further, our results on the transnational terrorism network provide some insight into how our approach will function on the often classified datasets of our target application. Currently, we are working with several organizations in the U.S. Department of Defense and the law enforcement communities to further study and transition this technology.

Next, in Section 2, we cover some technical preliminaries, including definitions of modularity. Section 3 describes the Louvain algorithm and our modifications to it to optimize for distance modularity. Section 4 describes our experiments and results on various data sets and an application to transnational terrorism. We review and place our work within related work in Section 5, and finally we conclude in Section 6. 

\section{Technical Preliminaries}

Throughout this paper, we shall model a network as an undirected graph $G=(V,E)$ where $V$ is a set of nodes and $E$ is a set of relationships among nodes.  We use $n,m$ to represent the cardninalities of $V,E$ respectively.  As the graph is undirected, we shall assume that $(v_i,v_j)\in E$ implies $(v_j,v_i)\in E$.  We also assume that each edge $(v_i,v_j)$ has an associated weight $w_{ij}$ (again $\forall i,j, w_{ij}=w_{ji}$).  For a given node $v_i \in V$, $\eta_i = \{v_j | (v_i, v_j)\in E \vee (v_j,v_i) \in E\}$ and $k_i = |\eta_i|$.

We shall use the notation $C= \{c_1,\ldots,c_q\}$ to denote a partition over set $V$ where each $c_i \in C$ is a subset of $V$, for any $c_i,c_j \in C$, $c_i \cap c_j = \emptyset$ and $\bigcup_i c_i = V$.

For a given partion, $C$, the modularity $M(C)$ is a number in $[-1,1]$ .  The modularity of a network partition can be used to measure the quality of its community structure. Originally introduced by Newman and Girvan.~\cite{newman04} this metric measures the density of edges within partitions compared to the density of edges between partitions.  A formal definition of this modularity (henceforth referred to as NG modularity) for an undirected network is

\begin{definition}[NG modularity~\cite{newman04}]
\label{regModDef}
Given partition $C= \{c_1,\ldots,c_q\}$, \textbf{NG modularity}, 
\[
M(C) = \dfrac 1 {2m} \sum_{c \in C}\sum_{i,j \in c}w_{ij}-P_{ij}
\]
where $P_{ij}=\frac{k_i k_j}{2m}$.
\end{definition}
Here, the null model used as a reference for comparison to a given partition assumes edges are rewired randomly,  while the degree sequence of the input network is preserved, hence $P_{ij}=\frac{k_i k_j}{2m}$.

Recently, a measure for modularity that accounts for distance, as well as network topology, was introduced by Liu et al.~\cite{liu12}.  Their modularity, henceforth referred to as distance modularity, is defined as follows:

\begin{definition}[Distance Modularity~\cite{liu12}]
\label{gsModDef}
Given partition $C= \{c_1,\ldots,c_q\}$, \textbf{distance modularity}, 
\[
M_{dist}(C) = \frac {1}{2m} \sum_{c \in C}\sum_{i,j \in c}w_{ij}-P_{ij}
\]
where $P_{ij}=\frac{\hat{P_{ij}}+\hat{P_{ji}}}{2}$, $\hat{P_{ij}}=\frac{k_i k_j f(d(v_i,v_j))}{\sum_{v_q \in V}k_q f(d(v_q,v_i))}$, and\\ $f : \Re ^+ \rightarrow (0,1]$ is the distance-decay function.
\end{definition}

The basic idea behind this distance modularity is that each node exerts a force on other nodes by generating a field, and the potential of the field at any point decreases with distance from the field source (the node generating the field), depending on the distance decay function \cite{li2007artificial,liu12}. The null model then that serves as a reference for comparison here  assumes that nodes which are closer according to the distance function are more likely to be connected. In this paper we shall assume the existence of a distance function $d:V \times V \rightarrow \Re^+$ that meets the normal axioms: $d(v_i,v_i)=0$, $d(v_i,v_j)=d(v_j, v_i)$, and $d(v_i,v_j) \leq d(v_i,v_q)+ d(v_q,v_j)$.

Previously, it has been proven that modularity-maximization is NP-hard~\cite{brandes08}.  Clearly, setting $\forall x, f(x)=1$, distance modularity reduces to NG modularity.  As a direct consequence of this observation, finding a partition that optimizes distance modularity is also NP-hard.

\begin{theorem}
Given graph $G=(V,E)$ and distance function $d:V \times V \rightarrow \Re^+$, finding a partition $C$ of $V$ that maximizes $M_S$ is NP-hard.
\end{theorem}

Throughout this paper, we will use an exponential distance-decay model\cite{liu12,cerina12,nekola99,skov01} defined as follows:
\[
f(x) = e^{-(x/\sigma)^2}
\]
Where $\sigma$ is a parameter in the interval $(0,\infty)$ and $e$ is the base of the natural logarithm.  One way to interpret $\sigma$ is that it is the distance where the force exerted by a point is reduced by a fraction $1/e$ (roughly $0.36$).  We note that in the limit as $\sigma$ approaches infinity, geospatial modularity reduces to NG modularity.  In the next section, we test a variety of settings for $\sigma$.  Learning parameters such as $\sigma$ has previously been explored in various geospatial applications -- see \cite{nekola99,skov01} for examples.

\section{Approach}

This section describes the approach we use to mine for geographically dispersed communities in networks. 
Although modularity maximization is NP-hard, a variety of practical approximation routines have been proposed~\cite{newman04,newman04a,blondel08} that experimentally have produced near-optimal partitions. 
In this paper, we employ the Louvain heuristic algorithm of Blondel et al.~\cite{blondel08}, only instead of using it to maximize NG-modularity, we use it to maximize distance modularity. In order to use the Louvain algorithm to maximize distance modularity, we must also modify some of it's steps. We summarize the Louvain algorithm briefly next (for more details on this algorithm, see~\cite{blondel08}) and describe our modifications and practical considerations when employing this heuristic algorithm to optimize distance modularity.

\subsection{Heuristic Algorithm}

The Louvain algorithm is an iterated, hierarchical process in which two phases are applied repeatedly until maximal modularity is reached: In the first phase, each node $v_i \in V$ of the given network is assigned to a community $c$, creating an \textit{initial partition}. In~\cite{blondel08}, the singleton partition was used. Then, for each $v_i \in V$, the gains in modularity that would result from placing $v_i$ to the community of each of its neighbors $v_j \in \eta_i$ are calculated, and $v_i$ is removed and placed into the community for which the maximum gain in modularity is attained (unless no positive gain in modularity is possible). This sub-process is repeated sequentially for each $v_i \in V$ until no individual move will result in a  gain in modularity, marking the end of the first phase and giving a partition $C$. 
In the second phase, a new network is built by using each $c_i \in C$ as a node in the new network, call these nodes \textit{meta-nodes}. Weights on the edges between any two meta-nodes in the new network are assigned to be the sum of the weights of the edges between nodes in the two communities corresponding to the meta-nodes. In this step, self-loops are created for each meta-node in the new network from the links between nodes of the community corresponding to that meta-node. After this phase is complete, the two phases are reapplied iteratively until there are no more changes. 

The efficiency of the Louvain algorithm relies on an easy re-calculation of modularity in the first phase of the algorithm.  When computing gains in modularity in phase one of the algorithm, removing any node $v_i$, the overall increase in modularity (regardless if it is distance or NG) if it is placed into community $c$ is proportional to:
\[
k_{i,in}-\sum_{j \in c}P_{ij}
\]
The only difference for distance modularity is that $P_{ij}$ is defined as per Definition~\ref{gsModDef} instead of Definition~\ref{regModDef}.
In terms of time complexity, the first phase of the algorithm is $O(n^{2})$, since for every node in the network, distance modularity must be computed according to Definition~\ref{gsModDef}, which is $O(n)$ in the denominator of $\hat{P_{ij}}$. The second phase is again $O(n)$. Both phases are a multiple of a constant that results from the number of iterations needed to run to completion. We note that the input sizes decrease drastically with each iteration, since communities are iteratively collapsed into nodes. Hence, the proposition on time complexity follows: 
\begin{proposition}
\label{quadProp}
The time complexity of the Louvain algorithm, optimizing for distance modularity, is quadratic in terms of the number of nodes $n$ of the input network. 
\end{proposition}

\subsection{Practical Considerations}

Apart from the main modification to use distance modularity instead of NG modularity, there are two steps of the original Louvain algorithm that we modify when optimizing for distance modularity. First, we must decide on an initial partition to use. Blondel et al.~\cite{blondel08} use the singleton partition. However, we have found that using the Louvain partition, resulting from a normal run of the Louvain algorithm optimizing for NG modularity, provided better results at the expense of some runtime. Second, since each node has an associated geospatial value, a geospatial value must be assigned to the meta-nodes of the new networks being built. Here we use the centroid of the communities that correspond to the meta-nodes. Throughout the remainder of this paper, we shall refer to the described modified version of the Louvain algorithm (for maximizing distance modularity) as the Louvain-D algorithm. The implications of these considerations are discussed in more detail in our experimental results.


\section{Experimental Results}
\label{expRes}

For our experiments, we used information extracted from the Gowalla and Brightkite location-based online social networking sites~\cite{cho11}.  

We built our implementation in Python 2.6 on top of the NetworkX library~\footnote{http://networkx.github.com/} leveraging code from Thomas Aynaud's implementation of the Louvain algorithm~\footnote{http://perso.crans.org/aynaud/communities/}.  Our implementation took approximately $1000$ lines of code.    The experiments were run on a computer equipped with an Intel X5677 Xeon Processor operating at 3.46 GHz with a 12 MB Cache running Red Hat Enterprise Linux version 6.1 and equipped with 70 GB of physical memory.  All statistics presented in this section were calculated using R 2.13.1.

\subsection{Distance Modularity Evaulation}
In our first set of tests, we iteratively selected nodes and their neighbors from the Brightkite network dataset provided by the authors of \cite{cho11} to produce $20$ small samples (of at least $300$ nodes each).  Each sample originated with a randomly selected node from the network and we iteratively added neighbors of the selected node(s) to the sample until a minimum desired sized was achieved.  Node and edge counts for these small networks is listed in Table 1.

\begin{table}
\label{smallNetData}
\caption{Brightkite Sample Data} \vspace{2mm}
\label{datasetTable}
\begin{center}
\begin{tabular}{|l||l|l|}
\hline
  & Nodes & Edges\\
\hline
\hline
Max& $331$ & $2801$   \\
\hline
Min& $300$ & $787$  \\
\hline
Avg & $311.25$ & $1560.40$  \\
\hline
\end{tabular}
\end{center}
\end{table}

On our $20$ samples extracted from the Brightkite dataset, we considered the straight-line distance between nodes in kilometers.  Hence, in calculating geomodularity, we ran experiments $\sigma = \{50, 100, 150, \ldots, 500\}$.  For each dataset and each value of $\sigma$, we compared the distance modularity returned by three approaches: the Louvain algorithm (which does not consider any geospatial information), the Louvain-D algorithm using singleton nodes as the initial partition, and the Louvain-D algorithm using the result of the Louvain algorithm as the initial partition.

\begin{figure}[htbb]
    \begin{center}
        \includegraphics[width=.95\linewidth]{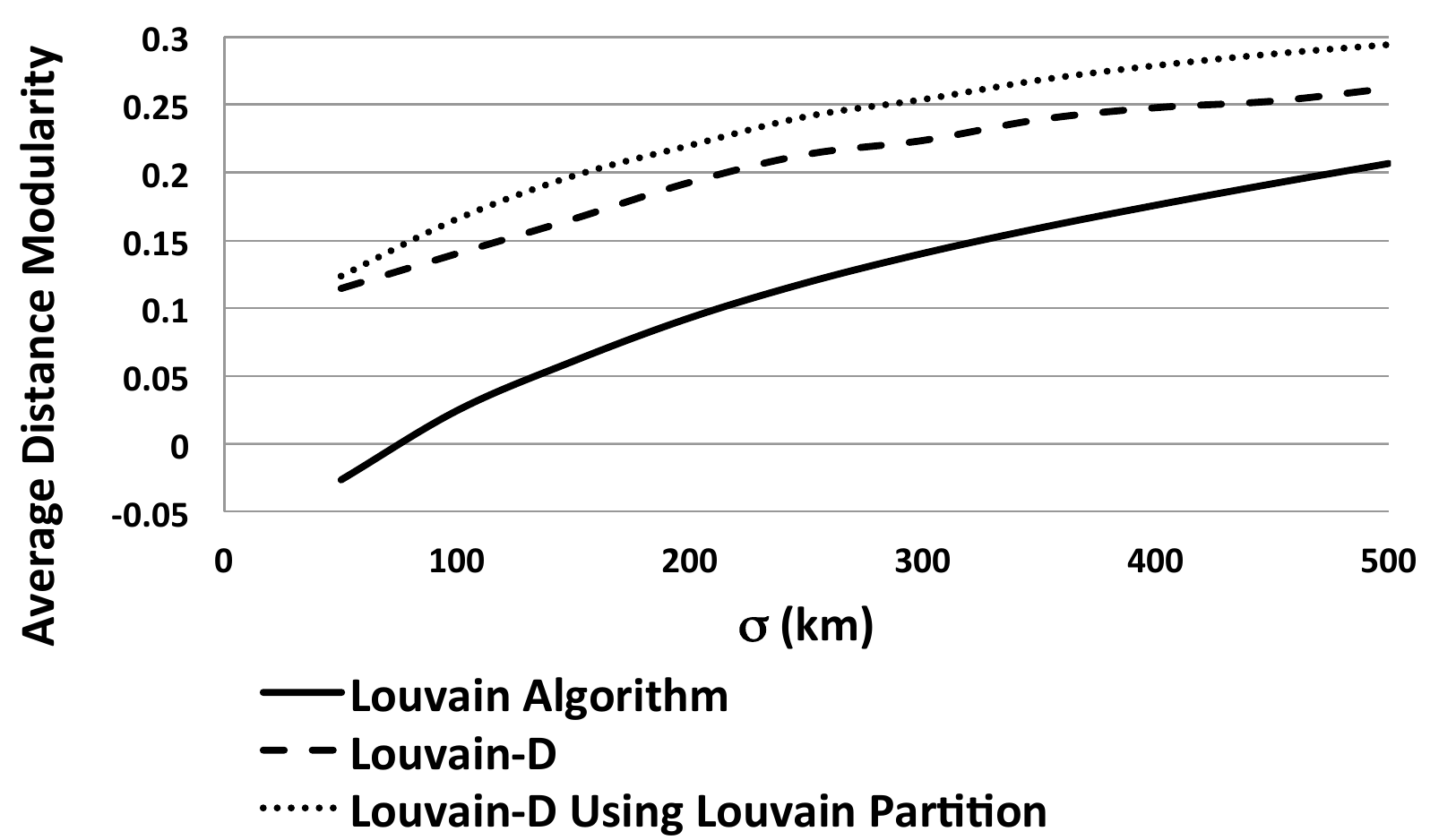}
    \end{center}
    \caption{$\sigma$ (in kilometers) vs. (average) distance modularity for the partitions returned by the Louvain-D and Louvain (baseline) algorithms.}
    \label{fig:avgGeomod}
\end{figure}

\begin{figure}[htbb]
    \begin{center}
        \includegraphics[width=.95\linewidth]{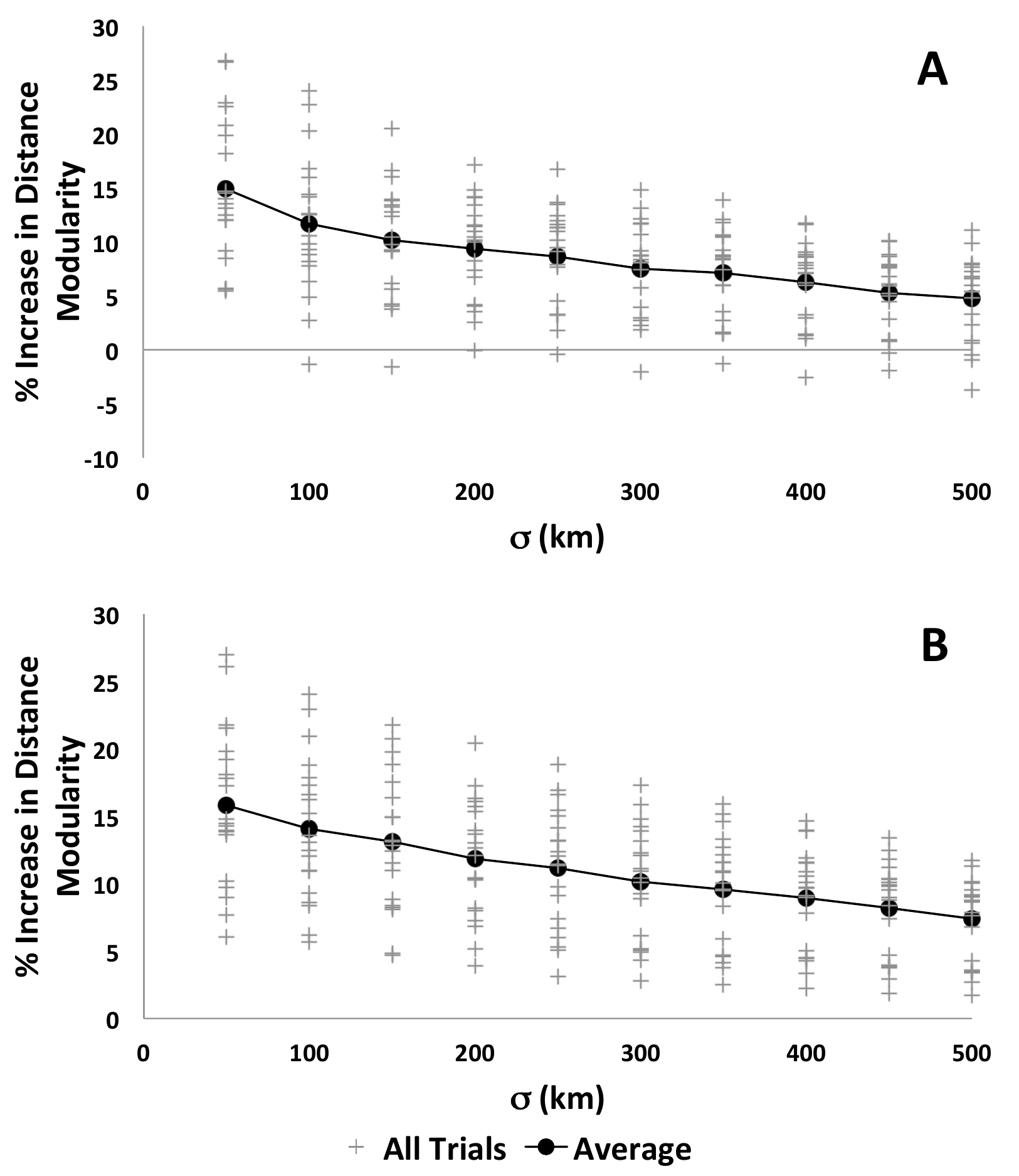}
    \end{center}
    \caption{$\sigma$ (in kilometers) vs. percent improvement in geomodularity (for the partition returned by the Louvain-D algorithm) when compared to the distance modularity for the partition returned by the Louvain (baseline) algorithm.  Panel \textbf{A} shows this relationship when the Louvain-D initially uses the singleton partition while panel \textbf{B} shows this relationship when the Louvain-D algorithm initially uses the Louvain partition.}
    \label{fig:percImrpove}
\end{figure}

Both variants of the Louvain-D algorithm returned a partition with a greater average geomodularity for each value of $\sigma$ than the partition returned by the Louvain algorithm (see Figure~\ref{fig:avgGeomod}).  This aligns well with the previous results of \cite{blondel11,cerina12} where space can affect on community structure not accounted for in the network topology.  However, we noticed that the percentage increase in modularity decreases with $\sigma$ (see Figure~~\ref{fig:percImrpove}).  This relationship also makes sense as distance modularity reduces to NG modularity (which the Louvain algorithm is designed to optimize) as $\sigma$ approaches infinity.

Although the Louvain-D algorithm outperformed the Louvain algorithm in terms of finding geomodularity, it generally returned higher-quality partitions if it was initialized with the Louvain partition instead of the partition of singleton nodes.  Further, when we used the Louvain partition to initialize the Louvain-D algorithm, we never obtained a partition with a lower geomodularity than the Louvain algorithm.  With the singleton partition, on the other hand, the Louvain-D was occasionally outperformed by the Louvain algorithm -- particularly for the higher values of $\sigma$.

\begin{figure}[htbb]
    \begin{center}
        \includegraphics[width=.95\linewidth]{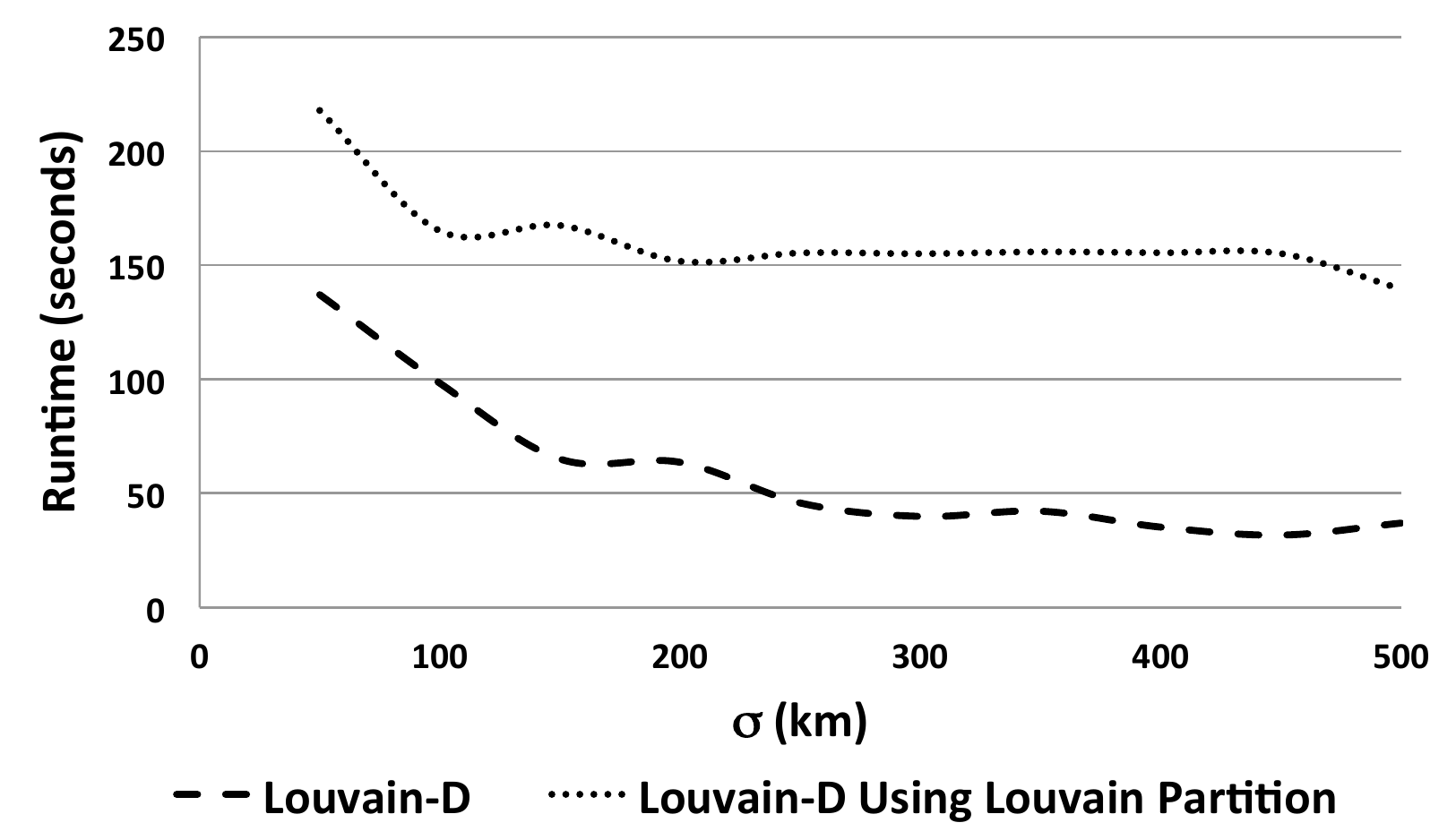}
    \end{center}
    \caption{$\sigma$ (in kilometers) vs. (average) runtime of the Louvain-D algorithm (using both singleton and Louvain partition initially).}
    \label{fig:runtime}
\end{figure}

However, the improvement in the quality when using the Louvain partition as a starting point comes at the expense of runtime.  While the time to calculate the Louvain partition was negligible (normally under $1$ second in the Brightkite tests), using it as a starting point appears to cause the Louvain-D algorithm to take longer to reach convergence - resulting in a runtime nearly double if the singleton partition is initially used (see Figure~\ref{fig:runtime}).

An analysis of variance (ANOVA) reveals that there is a significant difference in geomodualrity of the partitions returned by the three approaches on the Brightkite dataset ($p$-value $2.2 \cdot 10^{-16}$).  Additionally, pairwise analysis conducted using Tukey's Honest Significant Difference (HSD) test indicates that both instances of the Louvain-D algorithm provided results that differed significantly from the Louvain algorithm and each other with a probability approaching $1.0$ ($95\%$ confidence).  Additionally, the differences among runtimes were also significant (ANOVA $p$-value less than $2.2 \cdot 10^{-16}$) and pairwise different by the HSD with a probability approaching $1.0$ amongst all comparisons ($95\%$ confidence).
\pagebreak

\begin{figure}[htbb]
    \begin{center}
        \includegraphics[width=.95\linewidth]{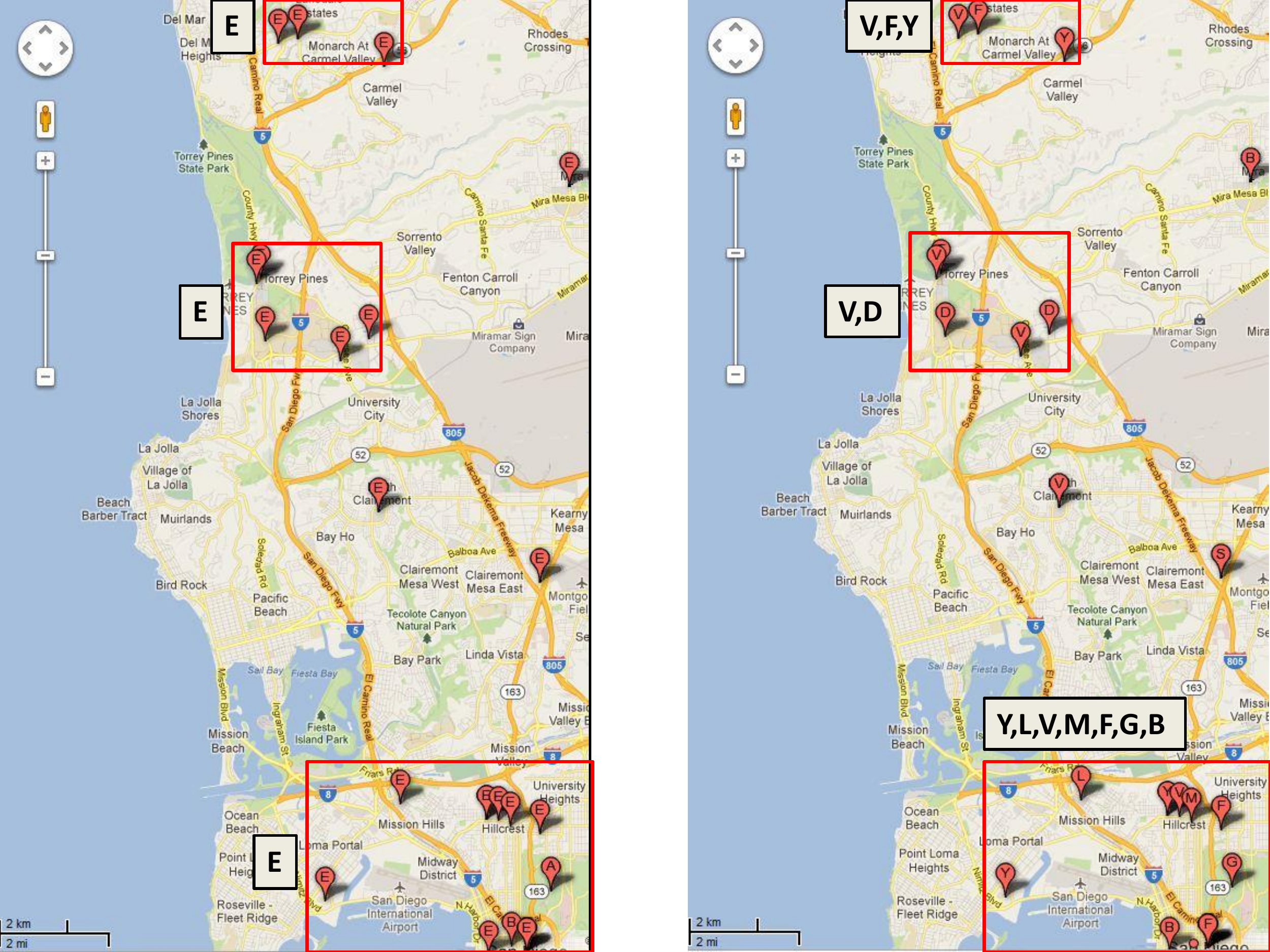}
    \end{center}
    \caption{\textbf{Left:} Communities identified using the Louvain algorithm, \textbf{Right:} Communities found using Louvain-D ($\sigma=150$)}
    \label{fig:sandiego}
\end{figure}

As an example of the type of result returned by our approach, we have included Figure~\ref{fig:sandiego} that illustrates the differences between a community returned by our approach vs. the standard Louvain algorithm.  The left panel shows a group of individuals near the San Diego area that the Louvain algorithm identified as being in the same community.  Likely, in this case, there is a strong correlation between geographic distance and connection in the social network.  The right panel, by contrast, shows that the same individuals are placed in multiple, different communities by the Louvain-D algorithm.  Since relatively high-degree individuals that are geographically near each other have a higher probability of connection in the null model, it becomes more likely for the Louvain-D algorithm to place them in different communities.

\FloatBarrier

\subsection{Tests on Larger Samples}

In our second set of tests, we iteratively selected nodes and their neighbors from the Gowala network dataset~\cite{cho11} to produce seven samples ranging in size from $301$ to $2101$ nodes each.  Samples were collected in the same manner as with the Brightkite samples previously described.  Distances between nodes are computed in kilometers.  Node and edge counts for these small networks is listed in Table 2.  Note that our tests examine networks significantly larger than those considered in related work where communities are determined based on geography and network topology ($100$ nodes in \cite{cerina12} and $571$ nodes in \cite{blondel11}).

\begin{table}
\label{lgNetData}
\caption{Gowalla Sample Data}\vspace{2mm}
\label{goDatasetTable}
\begin{center}
\begin{tabular}{|l||l|l|}
\hline
Sample No.  & Nodes & Edges\\
\hline
\hline
$1$& $301$ & $416$   \\
\hline
$2$& $602$ & $1550$   \\
\hline
$3$& $876$ & $12373$   \\
\hline
$4$& $1201$ & $2680$   \\
\hline
$5$& $1501$ & $3854$   \\
\hline
$6$& $1801$ & $4887$   \\
\hline
$7$& $2101$ & $6445$   \\
\hline
\end{tabular}
\end{center}
\end{table}

\begin{figure}[htbb]
    \begin{center}
        \includegraphics[width=.95\linewidth]{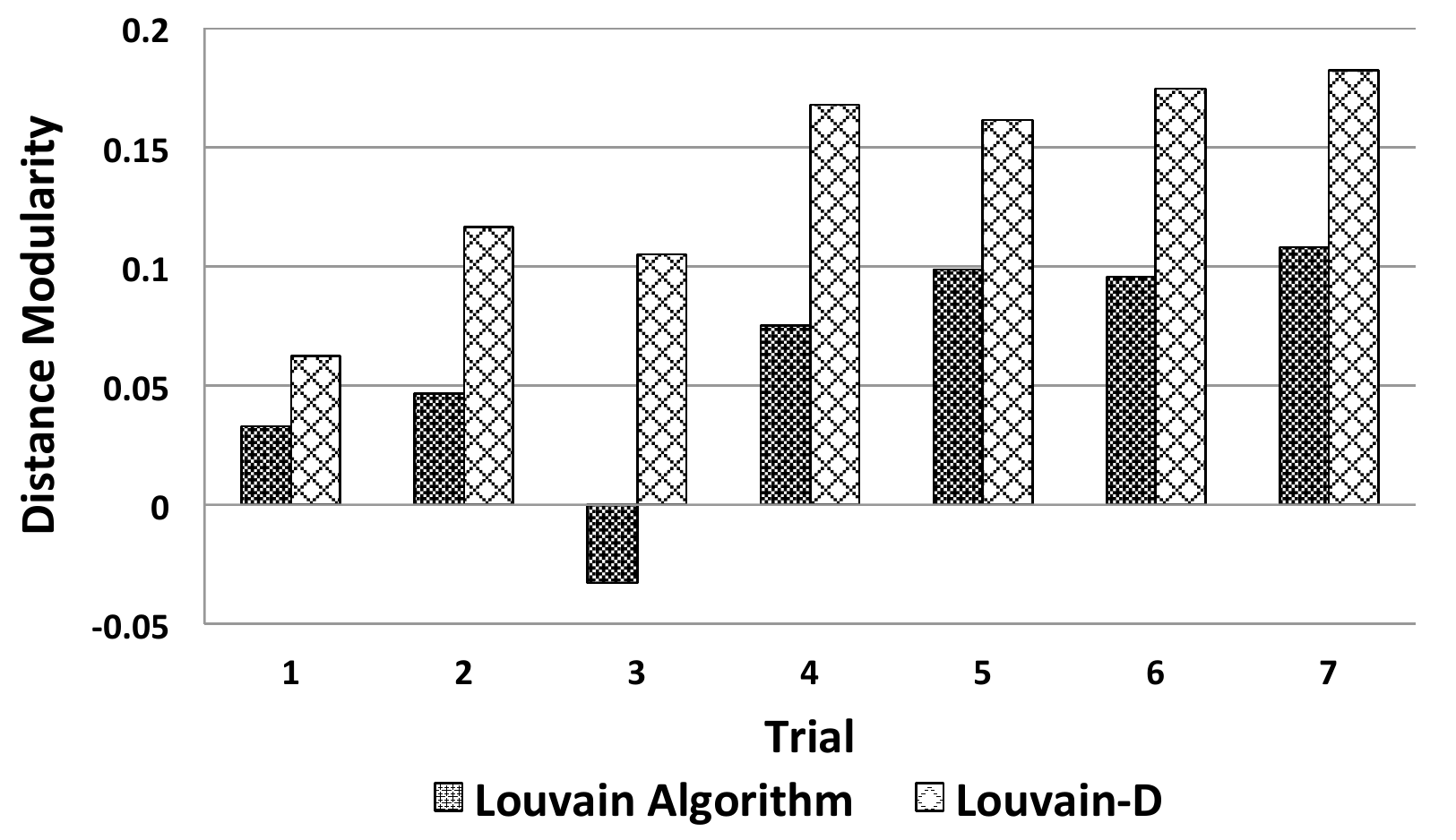}
    \end{center}
    \caption{Distance modularity of the partition found using the Louvain (baseline) and Louvain-D algorithms for the Gowalla network samples (see Table~\ref{goDatasetTable}).}
    \label{fig:goMod}
\end{figure}

We evaluated the Louvain-D algorithm on these samples with $\sigma=100$, initially using the Louvain partition, and compared the distance modularity of the resulting partition to that of the partition returned by the standard Louvain algorithm.  With all seven samples, the Louvain-D algorithm outperformed the standard approach.  Improvement ranged from $2.8$-$14.2\%$.  The results are depicted in Figure~\ref{fig:goMod}.

\begin{figure}[htbb]
    \begin{center}
        \includegraphics[width=.95\linewidth]{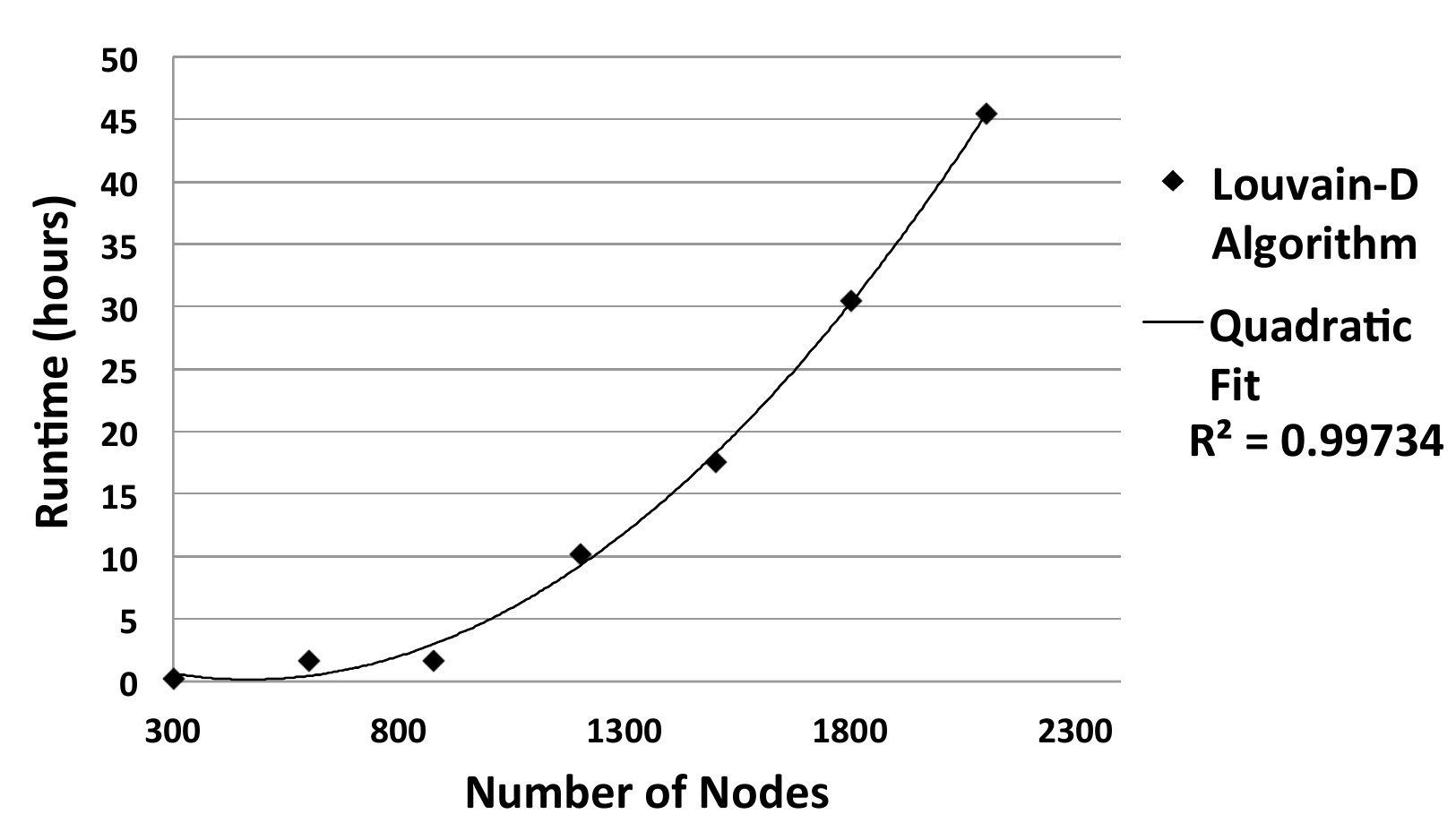}
    \end{center}
    \caption{Networks size (in nodes) vs. runtime (in hours) for the Gowalla network samples.  Note the strong quadratic fit.}
    \label{fig:goRuntime}
\end{figure}

We also studied the runtime of the Louvain-D algorithm and compared it to the size of the samples.  As per Proposition~\ref{quadProp}, we expected a quadratic relationship.  We verified this relationship in our experiment ($R^2=0.9973$).  These results are depicted in Figure~\ref{fig:goRuntime}.  We note that while considering a network of $2101$ nodes required just under two days of computer time, which is acceptable for our applications, further scaling will take prohibitively long runtimes.  For example, scaling to $10^4$ nodes would require approximately three months of runtime based on our regression analysis.  Further scalability is an important direction for future work.

\subsection{Application: Transnational Terrorism}

\begin{figure}[htbb]
    \begin{center}
        \includegraphics[width=.95\linewidth]{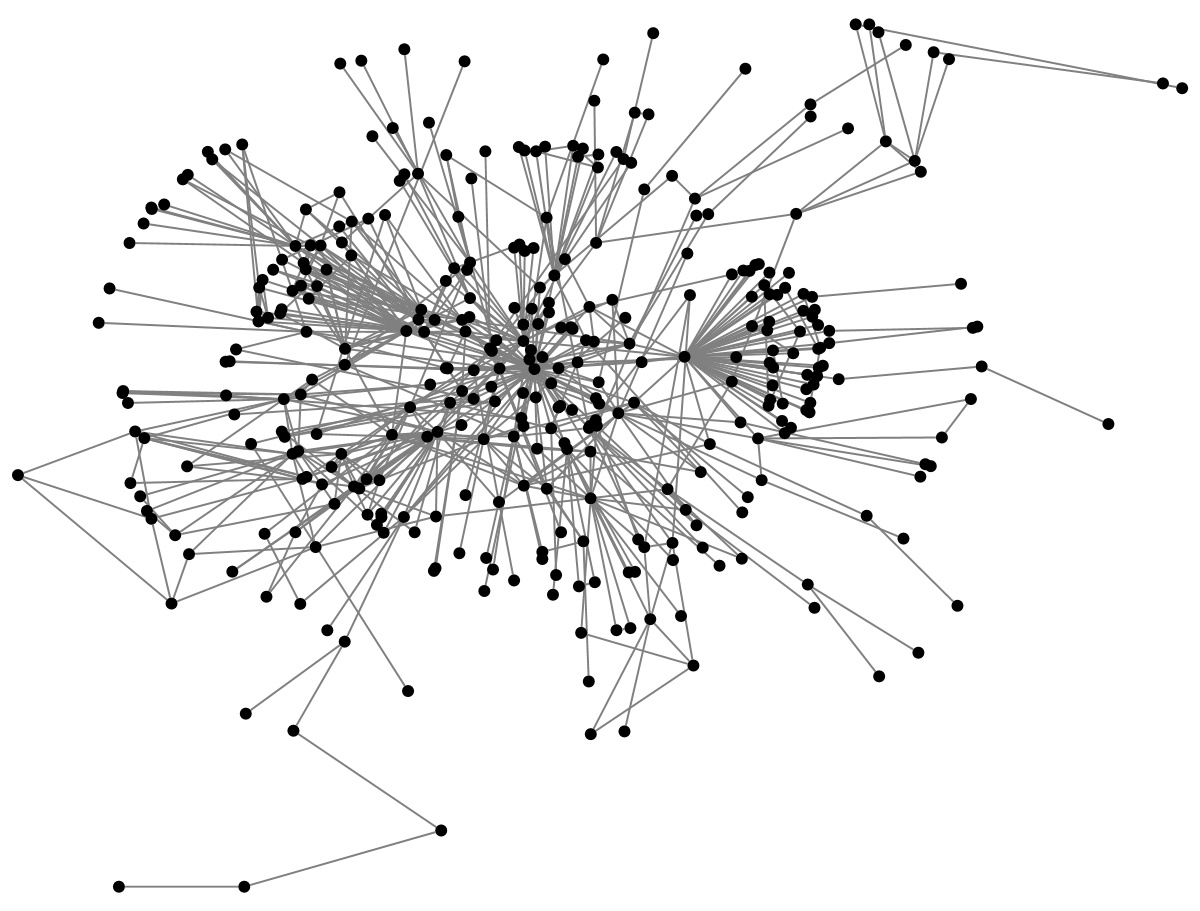}
    \end{center}
    \caption{Network relationships in the transnational terrorist dataset of \cite{medinaH11}.}
    \label{fig:fullTNW}
\end{figure}

\begin{figure}[htbb]
    \begin{center}
        \includegraphics[width=.95\linewidth]{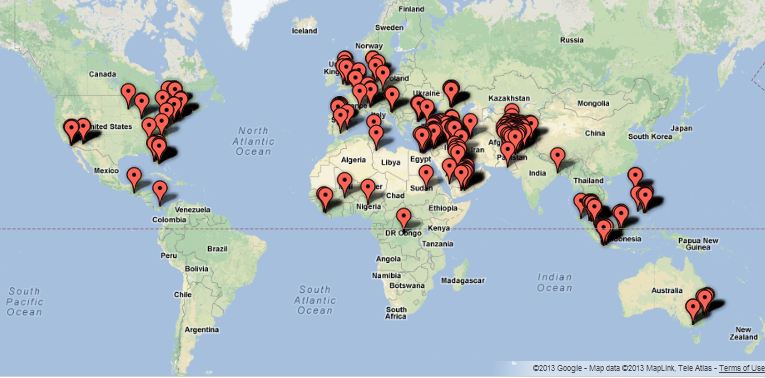}
    \end{center}
    \caption{Geographic locations of the individuals in the transnational terrorist dataset of \cite{medinaH11}.}
    \label{fig:fullTgeo}
\end{figure}

In this section we use the open-source derived terrorist network of Medina and Hepner~\cite{medinaH11} as a proxy for the (often classified) networks that will be used by this software in practice.  The networks consists of $358$ geolocated individuals in a transnational terrorist organization ($660$ unweighted edges).  A diagram of the network is shown in Figure~\ref{fig:fullTNW} while the locations of the individuals are shown in Figure~\ref{fig:fullTgeo}.

\begin{figure}[htbb]
    \begin{center}
           \includegraphics[width=.95\linewidth]{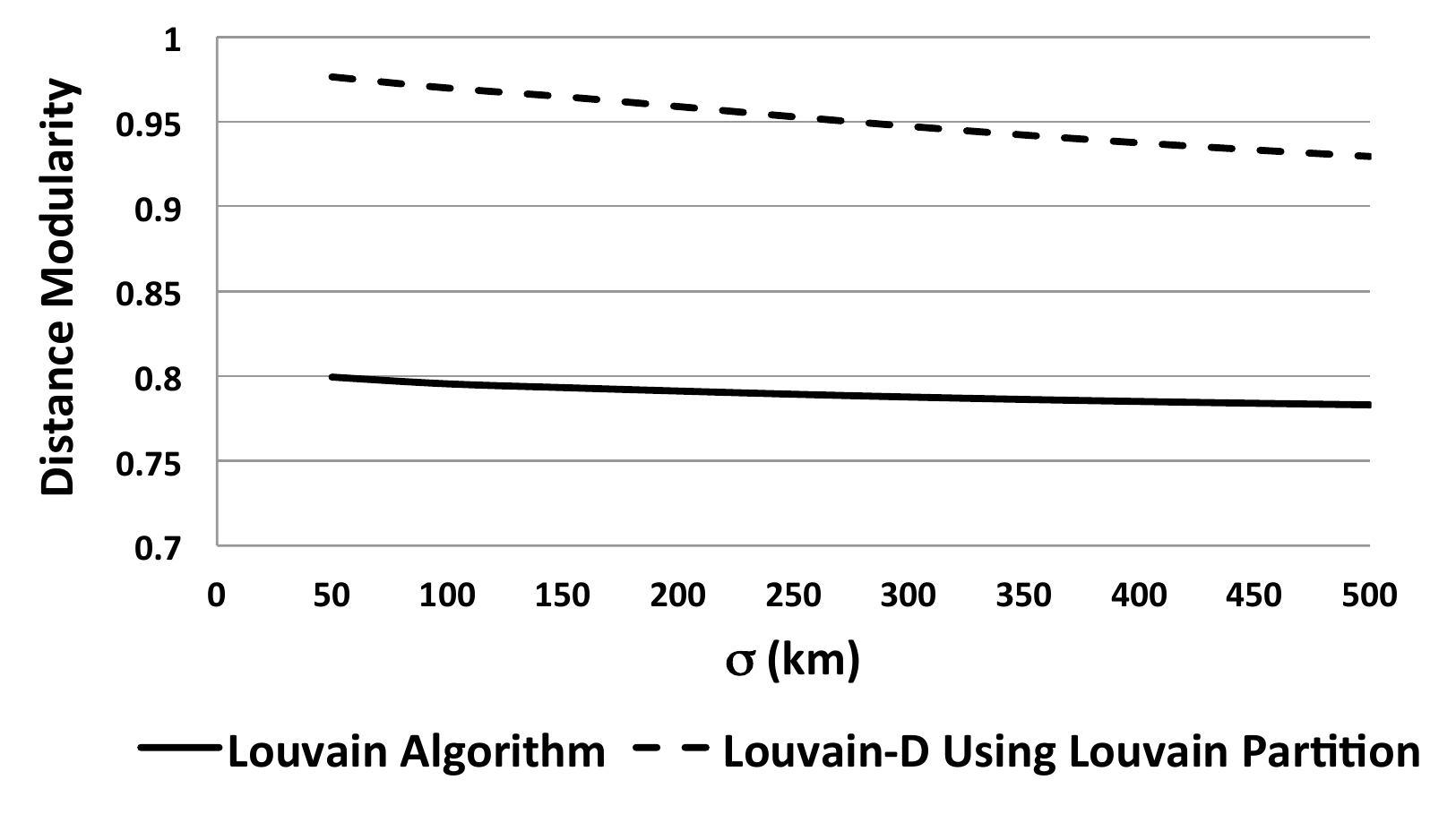}

    \end{center}
    \caption{Comparison of distance modularity between Louvain and Louvain-D algorithms for the transnational terrorism dataset.}
    \label{fig:topComGph}
\end{figure}

We ran the Louvain-D algorithm (initially using the Louvain partition) with $\sigma = \{50,100,150,\ldots,500\}$ and compared the distance modularity of the resulting partition to that returned by the standard Louvain algorithm.  The Louvain-D algorithm consistently outperformed the baseline approach (Figure~\ref{fig:topComGph}) with the percent improvement ranged from $8.2-9.8\%$.  The results are consistent with the other trials, where the distance modularity of the partition produced by the Louvain-D partition monotonically decreases with $\sigma$, slowly approaching the distance modularity of the baseline approach.

To better understand how a practitioner would use our approach for analysis, we considered the problem of identifying a single, important geographically disperse community.  We can identify such a group of individuals by determining the quality of a given community.  We can derive such a measure directly from the definition of modularity.  For a given given community $c \subseteq V$, we can determine the quality as follows:
\begin{equation}
M_c = \dfrac 1 {2 |c|} \sum_{v_i,v_j \in c}w_{ij}-P_{ij}
\end{equation}

\begin{figure}[htbb]
    \begin{center}
        \includegraphics[width=.95\linewidth]{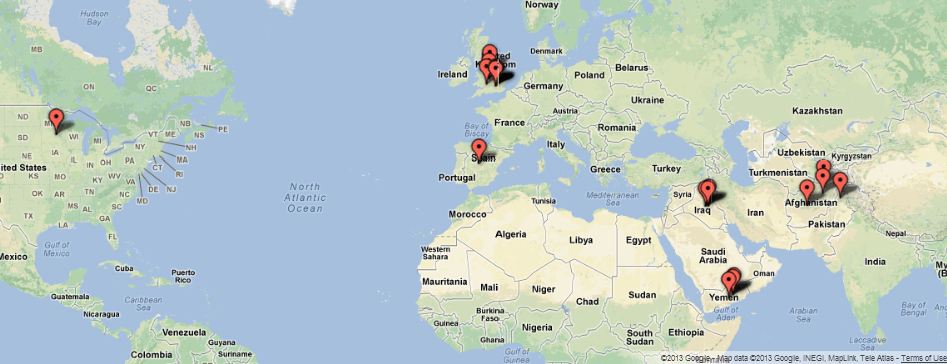}
    \end{center}
    \caption{Geolocations of the individuals in the top-ranked community from the transnational terrorist network.}
    \label{fig:topComTgeo}
\end{figure}

\begin{figure}[htbb]
    \begin{center}
        \includegraphics[width=.65\linewidth]{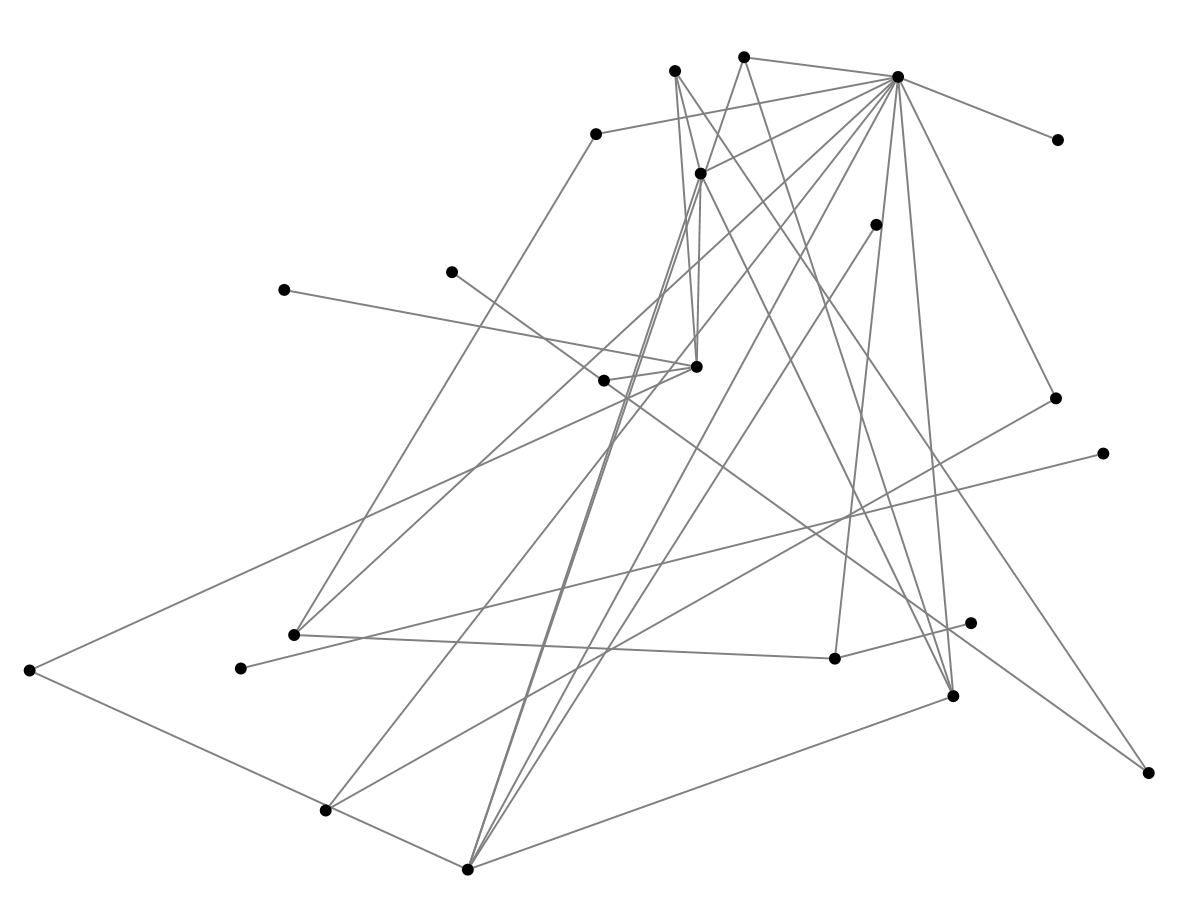}
    \end{center}
    \caption{Visualization of the network topology of the community shown in Figure~\ref{fig:topComTgeo}.}
    \label{fig:topComTNW}
\end{figure}

We ranked all the communities for the transnational\\ terrorist organization (over all settings of $\sigma$ we considered) and took the top one.  We show the visualization of the network and geolocations of the individuals in Figures~\ref{fig:topComTNW}-\ref{fig:topComTgeo}.  Note that the members of the identified community span three continents.  Identifying communities such as these can provide intelligence analysts insight into how various geographically-disperse terrorist cells interact with higher-level organizations.

\section{Related Work}
The use of modularity maximization for community finding was first introduced in \cite{newman04} which also described how to find partitions that could nearly maximize this quantity.  An exact method for addressing this optimization problem was introduced in \cite{brandes08}.  However, this method was based on integer programming and for many problem instances may take an exponential amount of time to complete.  However, we note that an easy modification of that program can be used to address the problem of this paper as the quantity $P_{ij}$ can be solved in a pre-processing step and treated as a constant in the integer program formulation.  Note that the time to complete such a step would be easily dominated by the overall runtime to even approximate a solution in such a method.  In the same paper, modularity maximization was also shown to be NP-hard, which precludes an efficient approaches under current theoretical assumptions.  In \cite{blondel08} the Louvain algorithm is introduced which is shown to provide partitions that nearly maximize modularity and can scale to very large networks.  The modification of the Louvain algorithm is what we leveraged in this paper.

Modularity was extended to consider geosptial relationships using a distance-decay model in \cite{liu12} with the introduction of distance modularity which we use in this paper.  Their approach modifies the null model to increase the expected number of edges between close nodes, it will tend to find communities that are more geographically disperse - hence meeting the requirement of our presented application.  Our work extends on their theory - providing an algorithm to find an approximately optimal partitions wrt distance modularity, experimental results, and describes practical considerations - none of which were included in \cite{liu12} which only introduces the the concept of distance modularity and describes the mathematical properties of their alternative null model.

The recent work of \cite{cerina12} introduces ``spatial modularity'' that also uses a distance-decay function in the null model - though somewhat different to that introduced in \cite{liu12}.  They study the difference among partitions created by attempting to optimize both standard modularity and their alternate definition on a series of small simulated networks whose edges are formed based on varying degrees of correlation between space and node similarity (determined by randomly assigned attributes).  The results of that paper have also inspired this work as they indicate that by considering geospatial relationships in the null model often yields different community structure than with the original definition of modularity introduced by \cite{newman04}.  However, unlike this paper, the work of \cite{cerina12} only studies simulated networks (this paper only looks at real-world networks).  The networks of this paper are an order of magnitude larger as \cite{cerina12} only considers networks of $100$ nodes.  Further, \cite{cerina12} does not describe any practical concerns in their approach that must be considered when creating a real-world system.  

Another important result on community finding in geosptial networks was that of \cite{blondel11} where the authors also modify modularity.  However, in that work, the authors use a null model that is based on an empirically observed probability distribution of edge existence based on distance.  Their optimization approach was tested on a network of Belgian communes of phone users and was shown to accurately identify linguistic communities.  However, unlike this paper and the work of \cite{liu12} and \cite{cerina12}, as their null model is based on an empirically determined probability distribution, it will not necessarily ensure geographically-disperse communities - which is our target application.  Further, the work of \cite{blondel11} does not describe practical considerations and their experimental evaluation is restricted to the Belgian phone network data consisting of $571$ nodes.

In addition to the aforementioned approaches, community detection in networks has also been explored in other manners that could potentially be proved applicable to geospatial applications - though to our knowledge no such application has been presented in the literature.  For instance, the work of \cite{yang09} identifies communities based on both network topology and content analysis.  Further, there are methods for community detection other than modularity maximization on networks (that do not consider spatial interactions).  Leveraging one of these other approaches is an important direction for future work. See \cite{fortunato09} for a comprehensive survey.

There has been other recent work where geospatial networks have been explored with respect to problems other than community finding.  The work of \cite{larusso12} discusses link-prediction and shows that by considering geography that results for this problem can be improved.  The work of \cite{abrol12} looks at identifying the location of users on Twitter using network topology.  Further, there also have been empirical studies on social networks with a spatial component such as \cite{barth11}.  Along such lines, the mobility of users in a location-based social network is explored in \cite{cho11,eagle2006reality}.  More domain-specific empirical studies related to this work are also prevalent in the literature.  Pertinent to our application include studies on terrorist networks~\cite{medinaH11} and criminal co-offender networks~\cite{Schaefer11}.

\section{Conclusion}

In this paper, we have presented a modified Louvain algorithm to find partitions of networks that provide near-optimal solutions for both nearness and distance modularity, providing a way to leverage spatial information in addition to network connection topology when mining networks for communities. We have evaluated this algorithm on two real-world location-based social networks, as well as a real-world transnational terrorism network data set. Our results have shown that using the Louvain algorithm modified to optimize for distance modularity to be an effective approach to the problem of finding geographically disperse communities, finding near-optimal solutions to distance modularity. Our experiments have also shown that using the Louvain partition instead of a singleton partition in the initial partitioning step of the algorithm generally provides improved final partitions in terms of distance modularity. We have demonstrated the scalability of the algorithm by considering networks of up to more than 2000 nodes, a number that is significantly greater than network sizes typically considered in the related literature. Finally, particularly through our experiments applying the algorithm to a real-world transnational terrorism network data set, we have found the presented approach be useful for finding geographically disperse communities at a time scale that is practical in the application domain.

Currently, examining scalability issues is an immediate concern for future work, as we have initiated a relationship with a major American police department to study gang violence - which will require the examination of networks of size $10^5$ nodes. In this application domain, the identification of particularly localized communities as opposed to disperse communities may be of interest as well, thus a modularity definition optimizing for this is another potential item for immediate future work. We are also working with various agencies in the U.S. Department of Defense to transition this technology to study networks of hundreds to thousands of nodes.  With this particular user-base, our focus is on readying the technology for deployment to analysts in a usable system.


\section{Acknowledgments}
P.S. would like to thank Richard M. Medina (GMU) for his help with the terrorist network dataset.  The authors are supported by the Army Research Office (project 2GDATXR042) and the Office of the Secretary of Defense.  The opinions in this paper are those of the authors and do not necessarily reflect the opinions of the funders, the U.S. Military Academy, or the U.S. Army.

%
%
%

\end{document}